\documentclass[aps,prl,twocolumn,showpacs,superscriptaddress,groupedaddress]{revtex4-1} 

\usepackage{graphicx,amssymb,amsmath,wasysym,color,overpic}
\usepackage{bm}  

\begin{document}

\title{Excess of Social Behavior Reduces the Capacity to Respond to
  Perturbations}

\author{David Mateo} \email[e-mail: ]{david.mateo.valderrama@gmail.com}\affiliation{Singapore University of Technology and
  Design, 8 Somapah Road, Singapore 487372} 

\author{Yoke Kong Kuan} \affiliation{Singapore University of Technology and
  Design, 8 Somapah Road, Singapore 487372}

\author{Roland Bouffanais} \affiliation{Singapore University of Technology and
  Design, 8 Somapah Road, Singapore 487372}

\begin{abstract}
Social interaction increases significantly the performance of a wide range of cooperative systems.
However, evidence that natural swarms limit the number of social connections suggests potentially detrimental consequences of excessive social activity.
Using a canonical model of collective motion, we find that the responsiveness of a swarm to local perturbations is reduced when the social interaction exceeds a certain threshold.
We uncover a similar effect for two distinct collective decision-making models of distributed consensus operating over a range of static networks.
While increasing the amount of interaction always increases the capacity of these systems to adapt to slow changes, an excess of social behavior can hinder the swiftness of their response to fast perturbations.
These results have far-reaching implications for the design of artificial swarms or interaction networks.
\end{abstract}

\maketitle

Social interaction is critical for swarms to perform an effective and
coordinated response to changing environments.  Social activity and the
associated transmission of information through the interaction network have
recently attracted considerable attention in a wide range of complex systems:
from the biological realm---flock of birds~\cite{attanasi14:_infor,Mor16}, school of
fish~\cite{sumpter08:_infor,%
  strandburg-peshkin13:_visual,herbert-read15:_initiat,calovi15:_collec},
swarm of insects~\cite{%
  attanasi14:_collec_behav_collec_order_wild_swarm_midges,%
  attanasi14:_finit_size_scalin_way_probe,Ni15}, and human
crowds~\cite{moussaid10:_walkin_behav_pedes_social_group}---and social
networks~\cite{fowler10:_cooper}, to artificial multi-agent systems such as the
power grid~\cite{alizadeh12:_deman_side_manag_smart_grid,%
  weckx14:_multiag_charg_elect_vehic_respec} and robotic
swarms~\cite{rubenstein14:_progr,kawashima14:_manip}.  The characteristics of
the interaction network are known to strongly affect the swarm
dynamics~\cite{young13:_starl_flock_networ_manag_uncer,%
  komareji13:_resil_contr_dynam_collec_behav,shang14:_influen} and, in
particular, its capacity to respond to local
perturbations~\cite{centola10:_spread_behav_onlin_social_networ_exper,%
  calovi15:_collec,%
  bassett12:_collec,%
  strandburg-peshkin13:_visual}.

Increasing the amount of social interaction usually improves the performance
of collectives, but it is known that most natural swarms operate with a
limited number of social connections.
For instance, flocking starlings interact on average with a fixed
number of conspecifics---6 to 7~\cite{cavagna}---and swarms of
midges~\cite{attanasi14:_finit_size_scalin_way_probe} regulate their
nearest-neighbor distance depending on the size of the swarm.  Gordon et
al.~\cite{gordon93:_what} have shown that one species of ants
(\emph{L. fuliginosus}) regulate its rate of social encounters following:
\emph{(i)} changes in the nestmate density for undisturbed ant colonies, and
\emph{(ii)} the introduction of an external perturbation---workers from
another colony---in the colony.  This limited interaction 
appears to be a behavioral feature and
not a direct result of physical limitations of their sensing
capabilities. These findings suggest that excessive social activity could be
detrimental to the collective dynamics, potentially
hindering its responsiveness to environmental changes.

Experimental evidence of such detrimental effects
has been found in the collective dynamics of
midges, where the
susceptibility of the system diminishes if the amount of
interaction---inferred from density---is increased above a certain value~\cite{attanasi14:_finit_size_scalin_way_probe}.
From the theoretical standpoint, some models of decision-making
dynamics predict that over-reliance on
social information can render a collective unresponsive to changing
circumstances~\cite{torney14:_social,kao14:_decis}. Models of consensus in mobile communicating
agents have also shown that consensus can be
reached more efficiently with a limited interaction range~\cite{baronchelli12:_consen}, which is strictly
equivalent to having a limited number of connections.
Understanding the consequences of excessive social activity is critical for achieving new functional predictions on collective animal behavior~\cite{strandburg-peshkin13:_visual,sumpter08:_infor}, and for the study of spreading of behaviors in networked systems such as online communities~\cite{centola10:_spread_behav_onlin_social_networ_exper}.
From a technological viewpoint, developing a predictive theoretical framework to understand under which circumstances these effects appear is of paramount importance for the emerging field of large-scale swarm robotics~\cite{rubenstein14:_progr,kawashima14:_manip}.

Here, we investigate the relationship between the number of social connections and the responsiveness of the collective.
First, we present an analysis of the correlations in swarms following a classical model of collective motion in which self-propelled particles (SPPs) move by adjusting their direction of travel to that of their neighbors.
Second, we extend the study beyond swarm dynamics by considering a general distributed decision-making model.
We measure how the properties of the interaction network affect the dynamics of the collective decision when facing external influences.
Finally, we develop an analytical framework based on linear time-invariant (LTI) theory that allows us to establish how the connectivity of a networked multi-agent system affects its overall responsiveness.
This framework may be used to determine policies on interaction regulation for optimal dynamical response to a given perturbation.

\section{Results}
\subsection[Correlations]{Correlations in swarms}
A natural starting point to characterize the responsiveness of the collective motion is to study the connected correlation in fluctuations of the consensus variable and the associated susceptibility~\cite{Ni15}.
It has been proposed that most biological systems may be poised near criticality~\cite{mora11:_are_biolog_system_poised_critic}, that is, they reside near the critical point between ordered and disordered phases where the system becomes highly correlated and thus has a large susceptibility.
This has been proven to be the case for swarms of midges lacking global collective order~\cite{attanasi14:_finit_size_scalin_way_probe}.
Specifically, it was observed that these swarms keep a low global alignment while exhibiting high levels of directional correlations~\cite{attanasi14:_collec_behav_collec_order_wild_swarm_midges,attanasi14:_finit_size_scalin_way_probe}. 

Being at the edge of chaos is apparently a favorable strategy for collectives faced with dynamical environmental perturbations.
However, some swarms such as starlings~\cite{Mor16} display a high degree of emergent global alignment.
If these swarms rely on a high level of order to perform collective actions---migration, milling, transport, etc---then being poised near criticality may not be a viable option for them.

\begin{figure}[htbp]
  \centering
  \includegraphics[width=0.98\linewidth]{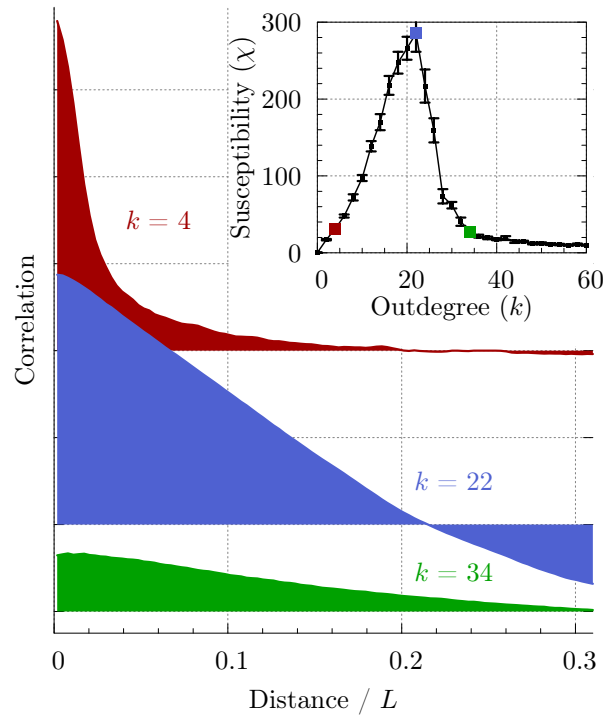}
  \caption{Trade-off in correlation length--strength. Correlation in velocity
    fluctuations (Eq.~\eqref{eq:correlation}) for $N=2,048$ topologically-interacting SPPs with outdegree
    $k=4, 22$ and $34$.  The distance is measured in units of the computation
    box length $L$.  Inset: Susceptibility (Eq.~\eqref{eq:susceptibility}) as a function of the number
    of neighbors or outdegree $k$. }\label{correlation_topo}
\end{figure}

Following the framework developed by Attanasi et al.~\cite{attanasi14:_collec_behav_collec_order_wild_swarm_midges,attanasi14:_finit_size_scalin_way_probe}, we compute the connected correlation in velocity fluctuations $C(r)$ (Eq.~\eqref{eq:correlation}) for a swarm composed of $N=2,048$ SPPs while varying the number of neighbors $k$.
We perform the calculations in the low noise regime
in order to investigate the behavior of swarms displaying a high degree of alignment~(see Methods).
Even in the ordered phase and far from the critical point, we find that the system can exhibit a large susceptibility if the amount of social interaction is set to an appropriate level (inset of  Fig.~\ref{correlation_topo}).
The susceptibility $\chi$ (Eq.~\eqref{eq:susceptibility}) is limited by the intrinsic trade-off between the spatial
spread of correlation and its short-range intensity: an increase in the
sociality (or amount of social interaction, see Methods) allows the information to travel farther through the
network---increased correlation length---but causes each agent to be exposed
to more information, thus decreasing the relevance of each individual
signal---decreased correlation strength.

The correlation function is shown in
Fig.~\ref{correlation_topo} for three different values of the number of
neighbors $k$, illustrating the clear trade-off between correlation spread
and intensity. For small values of $k$ (e.g. $k=4$ in Fig.~\ref{correlation_topo}), correlations are large but
confined to short distances.  As the amount of social interaction increases,
so does the spread of correlations and thus the susceptibility $\chi$.  Above
a certain optimal number of neighbors, which is approximately $k^* \simeq 20$ for the
particular configuration used in our calculation, the increase in spatial
spread cannot compensate the reduction in correlation strength and the
susceptibility of the system diminishes with increasing sociality.

\subsection[Predator Attack]{Collective response to a predator attack}

\begin{figure*}
  \centering
  \includegraphics[width=0.98\linewidth]{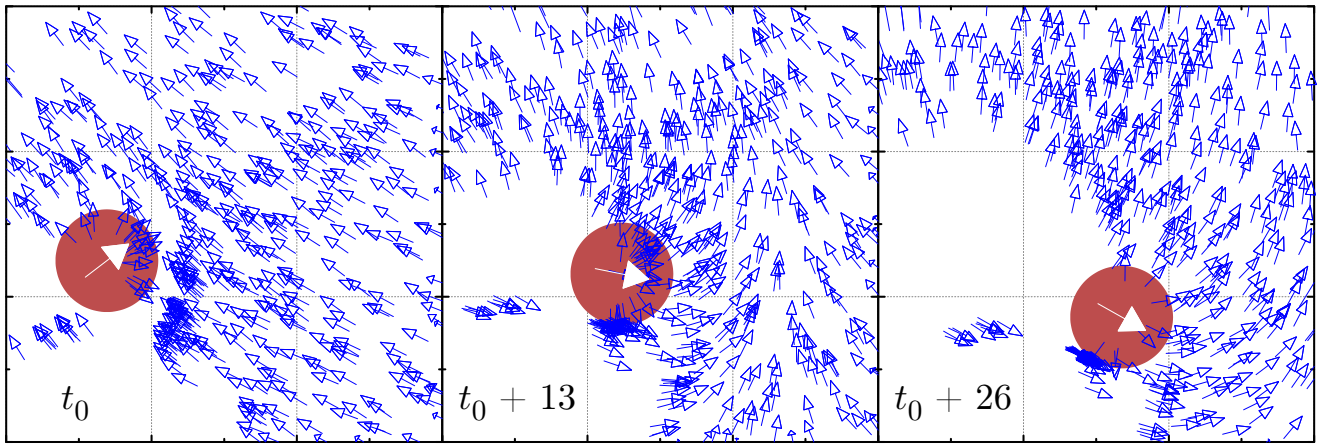}
  \caption{Collective evasive action induced by a predator attack.  The SPPs
    (empty blue arrows) can only detect the threat (solid white arrow)
    inside the danger-detection region shown in red.  In each consecutive
    frame, more agents outside this circle are able to flee without detecting
    the predator thanks to cooperative social behavior.  Each square in the
    background grid has a side of length $10\,\%$ that of the total
    computation box.  }\label{evasive}
\end{figure*}

In order to illustrate how a highly aligned swarm can benefit from a large susceptibility,  we have performed a model simulation of a predator attacking a group of SPPs following the Vicsek consensus and measured the survival rate of agents with different levels of social interaction (see Methods).
The emergent collective avoidance maneuver is shown in Fig.~\ref{evasive} for
three selected snapshots of a predator attack and in movies M1-M3 (see Supplementary Information).  At $t_0$ (leftmost frame), the predator starts
the attack on a highly-aligned section of the swarm.  Only the agents that
detect the predator---those inside the red disk---react according to
Eq.~\eqref{eq:flee}.  After 13 iterations, agents outside the detection area
are collectively reacting to the threat thanks to the social information
transmitted through the swarm.  After 26 iterations from the start of the
attack, all agents in the vicinity of the predator perform a global
evasive maneuver.
Notice that the information transfer has taken place strikingly fast,
which is in good agreement with recent empirical
observations of collective turns in flocks of starlings~\cite{
  attanasi14:_infor} and startled schools of
fish~\cite{herbert-read15:_initiat,rosenthal15}. 

It is worth pointing out that animals avoid predators using behaviors and strategies considerably more sophisticated than the model presented here.
For instance, Rosenthal et.~al~\cite{rosenthal15} have shown how schooling golden shiner fish use visual cues such as a fast change in speed to signal to other members the necessity to flee.
However, this idealized model illustrates how classical concepts from statistical mechanics such as the susceptibility can be linked to aspects of a collective's behavior crucial for the survival of its members.
A similar case could be made for the capacity of swarms to forage for food, achieve optimal pattern formation, or other examples of animal collective behavior where a timely response to perturbations is critical~\cite{Gel15,krause02:_livin_in_group}.

\begin{figure}[htbp]
  \centering
  \begin{overpic}[width=1.0\linewidth]{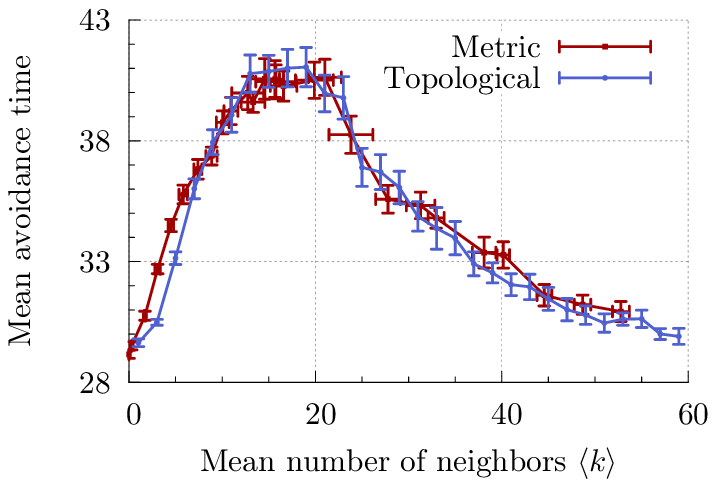}
    \put (20,55) {\large (a)}
  \end{overpic}
  \begin{overpic}[width=1.0\linewidth]{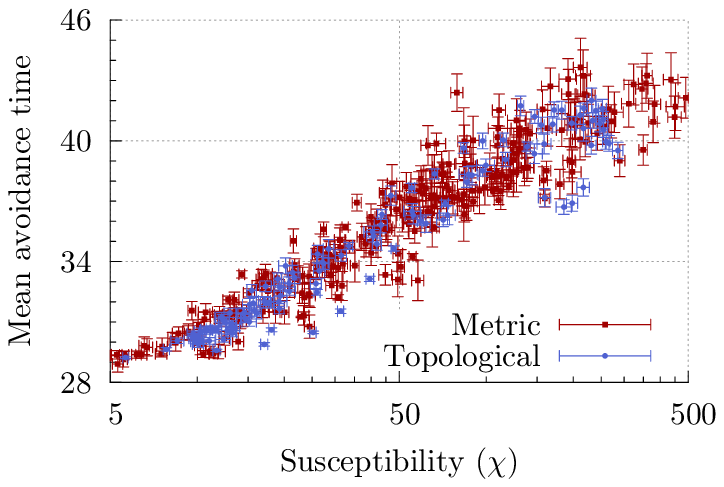}
    \put (20,55) {\large (b)}
  \end{overpic}
  \caption{Efficiency in predator avoidance. (a): Mean number of iterations
    between two consecutive predator catches (avoidance time) as a function of
    the average number of neighbors $\langle k \rangle$ for metric 
    (red, $\Box$) and topological (blue, $\ocircle$) interactions.  (b): Mean avoidance time
    as a function of the susceptibility $\chi$ of the equivalent unperturbed
    (in the absence of a predator) swarm.  }\label{survival_rate}
\end{figure}

The characteristic avoidance time for the swarm, defined as the average time
elapsed between two consecutive catches by the predator, is shown in
Fig.~\ref{survival_rate}(a) as a function of the mean number of neighbors
for both metric and topological interactions.  In the latter case, the mean
is exactly the imposed outdegree value $k$, while in the former
the average is computed over all agents and iterations. Interestingly, both
interactions yield essentially the same outcome.
Starting from a noninteracting collective ($\langle k \rangle=0$), the
avoidance time grows with the amount of social interaction up to a maximum
value about $40\,\%$ larger than the noninteracting time. From that optimal
point at approximately 20 neighbors, the avoidance time monotonously decreases
with increasing sociality, down to the value obtained for a
noninteracting collective.
In order to better understand how sociality influences the avoidance time, Movies M1 to M3 in the Supplemental Information present the movement of the swarm for each of the three characteristic regimes: optimal sociality ($k=16$), insufficient sociality ($k=4$), and excessive sociality ($k=40$) respectively.

Taking the avoidance time as a measure of the capacity of a swarm to respond
to localized perturbations, we can assess its relationship with
susceptibility.  Figure~\ref{survival_rate}(b) shows how the avoidance time
for a particular sociality varies with the susceptibility $\chi$ of the
equivalent unperturbed swarm.  Figure~\ref{survival_rate}(b) unambiguously shows a
systematic improvement in responsiveness with increased correlations.

\subsection{Influence of connectivity on collective decision-making}

We have seen that the susceptibility of an SPP system can be maximized by tuning social activity in the limit of low noise,
which may run against the intuition that the peak of susceptibility is located at the transition between the ordered and disordered phases~(see the Supplementary Note I for a discussion on this point).
This behavior is quite different from that of seemingly similar mechanistic models such as the $XY$ model in the highly ordered phase because the SPP is a nonpotential model with nonequilibrium dynamical effects~\cite{Ni15,toner95:_long,toner98:_flock}.

A model of collective behavior being nonpotential should not be surprising, as the agents or ``particles'' of the model are complex structures with the capacity of self-propulsion and, more importantly, an element of \emph{decision-making} that is absent in mechanistic models.
Thus, it is interesting to try to disentangle the effects of the collective decision-making process from the dynamics of the agents and the statistical description of the swarm.
To this aim, we turn our attention to a minimalist description of decision-making dynamics involving a set of fixed agents interacting through a static network and performing a consensus protocol.

An archetypical minimalist model of decision-making is the so-called linear threshold model, which is a generalization of the simple majority vote
model~\cite{aldana04:_phase}.
Using this model with different degrees of modularity,
Nematzadeh et al.~\cite{nematzadeh14:_optim_networ_modul_infor_diffus}
revealed that the network structure has a strong influence on information
diffusion.  A similar conclusion was obtained by
Centola~\cite{centola10:_spread_behav_onlin_social_networ_exper} using
experiments on a specifically designed social network. In both cases, the
effectiveness in information diffusion was characterized by the influence of
perturbations onto the asymptotic global state.

Here, we use the same model as Nematzadeh et
al.~\cite{nematzadeh14:_optim_networ_modul_infor_diffus} to study the
responsiveness of a decision-making process to perturbations. We characterize
this response capacity using the polarization speed $c$, which is essentially
the rate at which agents adapt their individual state to an induced
perturbations detected only by a minority of informed agents (see Methods).
These informed agents can be considered as ``leaders'' that drive the system
from $P=0$ to $P=1$, much like the SPPs detecting the predator
lead the swarm to perform a collective evasive maneuver or initiators can drive
sheep herds to specific targets \cite{Tou15}.

\begin{figure}[htbp]
  \includegraphics[width=1.0\linewidth]{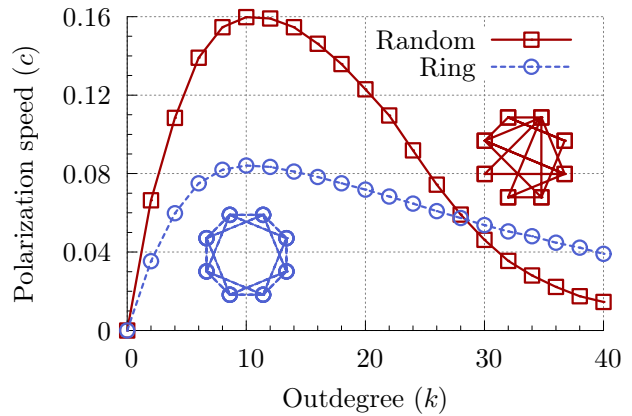}
  \caption{ Polarization speed $c$ for a linear threshold consensus protocol
    with threshold $\theta=1/2$ for a directed random network with fixed
    outdegree $k$ ($\Box$), and an undirected one-dimensional regular lattice
    with $k$-nearest neighbors connectivity ($\ocircle$).  The dynamics is
    triggered by switching $36\,\%$ of the $N=2,048$ agents to the state  $s=1$.
    }\label{voters}
\end{figure}

The polarization speed is shown in Fig.~\ref{voters} for two extreme kinds of
network wiring: a fixed-outdegree random directed network where each agent is randomly
connected with $k$ agents, and an undirected regular one-dimensional
lattice (a ring) where each agent is connected with its $k$ nearest neighbors.
With both wirings, the polarization speed $c$ is maximum for a finite
outdegree $k^*$ which, for large systems, is fairly independent of the total
number of agents $N$ (see Supplementary Note V).

We have chosen a large enough number of informed leaders to guarantee that the system eventually reaches $P=1$, i.e. a full polarization, in a finite time.
This way, the polarization speed is determined by the short-time response capacity of the system, and not its asymptotic polarization at long times.
How fast the system reaches full polarization depends on the amount of social interaction and,
as seen for the SPPs, too many connections hinder the performance of the system.

The results in Fig.~\ref{voters} also reveal that the structure of the network
can enhance or diminish the effects of connectivity on the response capacity.
While the optimal outdegree $k^*=10$ is the same for completely random and
highly structured networks, the polarization speed in the random network shows
a larger sensitivity to the amount of connections.
These results do not exhaustively prove the existence of such an optimal outdegree for any arbitrary topology. Nonetheless, it is
reasonable to assume that most realistic examples of complex networked systems
possess a network structure somewhere in between the two extreme cases
considered here~\cite{watts98:_collec}
(see Supplementary Note VI).
A systematic study of the polarization
speed for a wider collection of complex networks may reveal how the
short-time response of a system is related to other properties of the
interaction network such as degree distribution, average shortest connecting
path, and clustering coefficient~\cite{sekunda16:_inter}.

\subsection{Responsiveness of cooperative multi-agent systems}
In the previous sections, we have stressed the importance of distributed
consensus problems in both biological and social systems.
Both the SPP and the majority vote models provide excellent phenomenological frameworks to study
how the level of connectivity among agents affects the responsiveness of
cooperative systems.
However, their phenomenological nature limits our ability
to identify and characterize the underlying mechanisms responsible for the
impaired collective responses under excessive social connectivity.

Consensus and
cooperation in networked multi-agent systems is a topic that is starting to
receive significant attention in control theory and distributed computing
owing to numerous possible engineering
applications~\cite{olfati-saber07:_consen_cooper_networ_multi_agent_system}.
For instance, the power grid, urban traffic, arrays of distributed
sensors, multi-robot systems, and social networks are various examples of collective systems
requiring an effective response to local perturbations. 
The design of such systems---especially in the emerging field of swarm robotics---can be optimized using a theoretical framework that highlights the underlying mechanism and predicts under which conditions the detrimental effect of excessive connectivity will manifest.
The LTI system theory provides one of the most elementary candidates for such a framework.

We consider a set of $N+1$ agents performing a linear consensus protocol, and
model the effects of local perturbations by setting one agent as a ``leader''
that does not participate to the local consensus dynamics albeit influencing
agents connected to it.
The leader-follower distributed linear consensus
protocol presented in Eq.~\eqref{eq:local-consensus} is fairly
standard~\cite{%
  olfati-saber07:_consen_cooper_networ_multi_agent_system,%
  jad,young13:_starl_flock_networ_manag_uncer,komareji13:_resil_contr_dynam_collec_behav,%
  shang14:_influen,shang14:_consen},
and can be used to analyze the capacity of the system to follow and adapt to fast changes
in the behavior of the leading agent (see Methods).
To simplify the problem as much as
possible, static and undirected regular one-dimensional lattice
topologies---$k$ nearest neighbors with a ring topology---are considered for
the network of interaction between agents.

Significant attention has been dedicated to the problem of convergence to
consensus~\cite{shang14:_influen} and controllability of multi-agent
dynamics~\cite{komareji13:_contr} in the presence of complex network
topologies---possibly switching---with directed or undirected information
flow~\cite{%
  olfati-saber07:_consen_cooper_networ_multi_agent_system}.  Here, given the
simple topology of the static network, both convergence to consensus and
controllability are guaranteed.
Instead, our focus lies with the overall responsiveness of the collective in
adapting to fast changes in the dynamics of the single leader---in
control-theoretic terms, the input.

\begin{figure}[htbp]
  \includegraphics[width=1.00\linewidth]{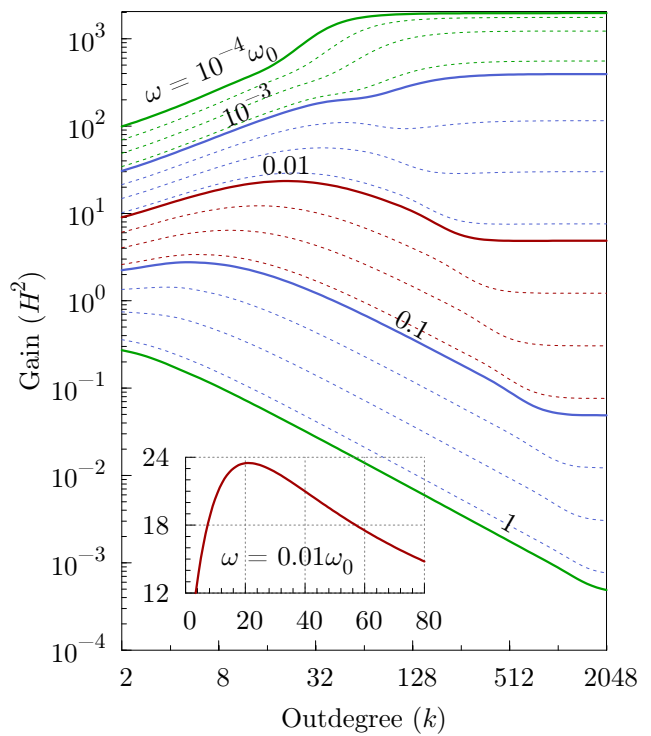}
  \caption{Responsiveness of a distributed consensus leader-follower protocol.
    The total amplitude gain for a system of $N=2,048$ agents following a single
    leader as a function of the outdegree $k$ (number of connections) is shown for
    several values of the leader's oscillating frequency $\omega$.
    Each solid line corresponds to a frequency equal to the agents' natural frequency $\omega_0$ times the factor specified on top of the line.
    The dotted lines correspond to frequencies 2, 4, and 8 times larger than the one in the solid line above them. 
    Inset: detail view of the gain for $\omega=0.01\omega_0$ in the region where the gain presents a maximum with respect to the outdegree.
    }\label{state-space}
\end{figure}

To characterize the effects of varying levels of connectivity (or sociality)
on the far-from-consensus responsiveness of the collective,
Fig.~\ref{state-space} shows the response capacity of this system to
oscillations of the leading agent as a function of the number of connections
$k$,  and for input
oscillation frequencies $\omega$ spanning four orders of magnitude.
This response capacity, measured by the total amplitude gain (see Methods), can be roughly interpreted as the number of agents that are capable of following the perturbation induced by the leader.

For such a rudimentary linear system, the
response capacity  exhibits a surprisingly rich structure.
At low frequencies $\omega\ll\omega_0$, an
increase in sociality always translates into an improvement in the system's
capacity to respond to perturbations. At high frequencies
$\omega\gtrsim\omega_0$, the opposite is true: adding connections
systematically yields a reduction in the system's performance. A very interesting
intermediate frequency regime is also observed (e.g. $\omega\sim 0.01\omega_0$ in Fig.~\ref{state-space}), where the responsiveness features a peak at a finite level of sociality.
The inset highlights that, in this intermediate frequency regime, the system can essentially double the amount of agents capable of following the leader by tuning the outdegree to its optimal value.
This trend is reminiscent of what we have uncovered for the variations of the
susceptibility (Fig.~\ref{correlation_topo}), mean avoidance time
(Fig.~\ref{survival_rate}(a)), and polarization speed (Fig.~\ref{voters}) as
a function of the sociality $k$ in the previous phenomenological models.

Using the analytical expression for the gain,
we can obtain general predictions for the responsiveness on arbitrary networks in the limits of low and high frequencies.
For instance, in the limit of low frequencies, any system with an undirected interaction network (for directed ones, see Supplementary Note VIII) will respond to perturbations of frequency $\omega\ll\omega_0$ as
\begin{equation}
H^2_{\omega\ll\omega_0} = \|{\mathbf H}_0\|^2  - \omega^2 {\mathbf H}_0^\dagger{\mathbf W}^{-2}{\mathbf H}_0 + O(\omega^4),
\label{eq:limit_low}
\end{equation}
where ${\mathbf H}_0={\mathbf H}(0)={\mathbf W}^{-1}{\mathbf W}_l$ is the gain at $\omega=0$, ${\mathbf W}$ is related to the inter-agent connectivity, and ${\mathbf W}_l$ to the connectivity of the agents to the leader (see Methods).
For any connected network,  ${\mathbf H}_0$ is a vector with all components equal to 1~\cite{jadbabaie03}, and thus Eq.~\eqref{eq:limit_low} can be written as
\begin{equation}
H^2_{\omega\ll\omega_0} = N  - \omega^2 \sum_{i,j}({\mathbf W}^{-2})_{ij} + O(\omega^4).
\label{eq:limit_low2}
\end{equation}
At high frequency, the gain is
\begin{equation}
H^2_{\omega\gg\omega_0} = \omega^{-2} \|{\mathbf W}_l\|^2 + O(\omega^{-4}).
\label{eq:limit_high}
\end{equation}

Note that the low-frequency limit is fully determined by the connectivity between agents ${\mathbf W}$, while the high-frequency limit only depends on the connectivity of the agents with the leader ${\mathbf W}_l$.
For networks with a fixed  outdegree $k$, the norm of this connectivity decreases as $\|{\mathbf W}_l\|^2\sim 1/k$.
Thus, at high frequencies, distributed consensus systems have a decrease in responsiveness with increasing number of connections.
This is a general behavior and not a particular feature of the regular ring network used in the previous calculations.

In general, there is no such direct relation between ${\mathbf W}^{-2}$ and the amount of connections in the network, which means that the behavior of the system at low frequencies is more sensitive to the features of the interaction network beyond its outdegree distribution.

From the standpoint of designing artificial swarms, this
analysis highlights that the pace of typical perturbations faced by the system is central in defining appropriate levels of interagent
connectivity.  When subjected to slow-changing perturbations, the system's
effectiveness always benefits from a higher level of connectivity. Comparing
with earlier observations, one can deduce that in the low-frequency regime,
the system does not require high correlation strengths for good propagation of
the signal, but it does benefit from an increase in speed that higher
correlation lengths provide.  On the other hand, fast perturbations
inevitably reduce the system's effectiveness with increasing interagent
connectivity.  Extending the comparison, this suggests that in the high-frequency
regime, a high correlation strength is paramount for the signal to
be effectively transmitted throughout the entire system.

\section{Discussion}
A myriad of organisms manifests swarming and social organization to
some degree.  It is well known that such collective behaviors notably improve
the effectiveness of fundamental tasks, e.g. predator avoidance, foraging, or
mating. However, our phenomenological and analytical study of different models of collective behavior 
reveals that an excess in social interaction can have detrimental effects, in
that it leads to a reduced capacity of response to fast localized perturbations.

Specifically, we have shown that for a system of self-propelled
agents---subjected to a consensus protocol to align their velocities---the
susceptibility of the swarm is maximized for a finite amount of social
interaction. In other words, the responsiveness of the swarm is reduced if the
sociality is increased above a certain level.
Beyond the field of swarming, we have
found that simulations of a minimalist model of collective
decision-making---the linear threshold model---exhibits a reduced capacity to respond to perturbations with excessive sociality.
Lastly, a frequency-domain analysis within the LTI framework reveals the underlying cause of this phenomenology: in general, more social interaction in a multi-agent system increases its responsiveness to slow perturbations while decreasing its responsiveness to fast ones.

Simulation of an idealized predator attack upon a swarm of SPPs following the Vicsek model with both metric and topological interactions reveal a direct
connection between the high susceptibility $\chi$ of the swarm and the
survivability of its members in hostile environments.
We speculate that the dilution of information transferred through the swarm occurring for high levels of sociality may be the reason behind the apparently self-imposed limit on social
activity observed in flocking birds~\cite{cavagna}, social
ants~\cite{gordon93:_what}, and other taxa.

In terms of correlations, an increase in the number of neighbors yields an increase in the
correlation length at the cost of decreasing the correlation strength.
At low sociality (e.g. below $k=20$ in Fig.~\ref{correlation_topo})
this is a beneficial trade-off for the swarm: the increase in correlation
length effectively allows the information to propagate faster through the
interaction network. Thus, more agents are capable of responding to the
presence of the threat. However, at high sociality
(e.g. beyond $k=20$ in Fig.~\ref{correlation_topo}), the increase in
correlation length only affects agents far away from any danger and marginally
benefits the overall performance of the swarm. On the other hand, this
increase in the correlation length is accompanied by a drastic reduction in
correlation strength that, in turn, severely reduces the responsiveness of
agents in the vicinity of the threat.

Sociality has similar effects on responsiveness for the case of multi-agent
systems performing distributed consensus with a threshold-triggered dynamics,
meaning that an agent only changes its state when a certain amount of its
neighbors do.
These kind of threshold events have been observed in the spread of information
over social networks \cite{centola10:_spread_behav_onlin_social_networ_exper} and the so-called flash expansion of whirligig beetles
facing a potential predator threat~\cite{Rom15}.

It is worth pointing out that our results for dynamical responsiveness
complement previous studies associated with global properties, such as the
robustness of the interaction
network~\cite{young13:_starl_flock_networ_manag_uncer} or the consensus
speed~\cite{shang14:_influen}. In these studies, increasing the amount of
interaction eventually yields diminishing returns---i.e. less gain per
neighbor---but never an actual reduction in the property of interest.
Diminishing returns can only justify the preference for a finite number of
connections if the cost for establishing links between agents is
significant. However, quantifying such costs in biological
swarms is close to impossible given the
complexity associated with sensory and neurological requirements~\cite{young13:_starl_flock_networ_manag_uncer,shang14:_influen}.
In contrast, the present study on the dynamical responsiveness of the swarm shows
an absolute reduction in swarming effectiveness when the number of neighbors
is increased above a certain level.

This fact raises the important question of why collectives with excessive
connectivity display a reduced effectiveness under some scenarios such as a
predator attack, but not under others such as consensus reaching.
The present analysis of the responsiveness of multi-agent systems
following LTI consensus dynamics under time-varying perturbations reveals that
one key element for predicting the effect of connectivity on responsiveness is
the speed of perturbation changes.
In many cases, being able to react efficiently to perturbations in the appropriate time scale is essential for the performance of systems conducting distributed consensus.
For example, ants performing collective transport of food rely on transiently informed ants to locate their nest~\cite{Gel15}.
These informed ``leaders'' forget their knowledge after a time of joining the collective action, and thus provide a changing signal with a certain characteristic time scale to the swarm. 
Successful transport depends both on a high consensus over the direction of movement and a proper responsiveness to this dynamic input.

As can be seen in Fig.~\ref{state-space}, high
levels of connectivity provide marginal benefits when the system is subjected
to slow perturbations, but yield a sizable reduction in effectiveness in the
presence of relatively fast perturbations.

In summary, previous studies in the animal realm \cite{cavagna, attanasi14:_finit_size_scalin_way_probe, gordon93:_what} and in social systems \cite{centola10:_spread_behav_onlin_social_networ_exper} provide evidence suggesting that, in some cases, it is optimal for collectives to limit the amount of social interaction.
We have presented a statistical treatment of a standard SPP model for collective motion, simulations of decision-making dynamics, and an analysis of the frequency-response in a consensus protocol that consistently exhibit a decreased responsiveness associated with an excess of connection or interaction.
Given that these models are relatively general and unadorned, we suggest that this non-trivial relation between responsiveness and sociality may be a general feature of a wide
range of complex systems involving distributed consensus.

Besides shedding a new light on our understanding of
collective behavior, this has also clear implications for the design of
networked systems. Even ignoring the possible costs of establishing
connections and transmitting information between agents, it may be desirable
to limit the number of connections in order to achieve a more effective dynamical response.

\section{Methods}

\subsection{Self-propelled particles}

We use the self-propelled particles (SPP) model developed by Vicsek
et al.~\cite{vicsek95:_novel} as a minimalist model of collective motion that
captures the cooperative alignment of orientation. There are several
extensions and improvements to this model that generate more realistic and
specific dynamics~\cite{viscek2012}, but we use the original model for the
sake of generality and simplicity. Each particle moves in a two-dimensional
periodic space and changes its direction of motion at discrete timesteps in
order to align to its neighbors' mean orientation according to
\begin{align}
  \mathbf{x}_i(t + \Delta t) &= \mathbf{x}_i(t) + \Delta t \,\mathbf{v}_i(t) ,\nonumber \\
  \theta_i(t + \Delta t) &= \arg\Big(\sum_{j\sim i} \mathbf{v}_j(t)\Big) +
  2\pi\eta_i(t),
  \label{vicsek}
\end{align}
where the velocity vector $\mathbf{v}_i=v_0\bm{\hat{\theta}}_i$ has constant magnitude
$v_0$ and direction $\theta_i$, $\arg()$ gives the orientation of a vector, and $\eta_i(t)$ is a random number uniformly
distributed in the $[-\eta/2,\eta/2]$ range.  The sum $j\sim i$ is performed
over the neighbors of $i$ (including $i$ itself).

While the original Vicsek
model considers that a pair of agents interact---i.e., are neighbors---if they
are closer than a certain distance (metric interaction), there is strong
evidence that certain natural systems such as flocks of birds interact with a
fixed number of neighbors instead (topological or metric-free
interaction)~\cite{cavagna,ginelli10:_relev_metric_inter_flock_phenom}. For
this reason, we have studied different kinds of interactions only to find the
same phenomenology; the responsiveness depends essentially on the amount of
interaction in the swarm, not the details of the interaction rule itself.
Thus, throughout this work, we talk about \emph{sociality} to refer to the
parameter quantifying the amount of interaction between agents, be it the
interaction radius in the metric case or the outdegree (number of neighbors)
in the topological one.

The results presented in this work have been obtained by computing the
dynamics of a set of $N=2,048$ SPPs following the Vicsek model starting from
random positions and velocity orientations.  The numerical calculations have
been performed using the \texttt{libspp} library~\cite{swarming-spp}.
Further implementation details can be found in Supplementary Note V.

\subsection{Correlations and susceptibility}

%
The dimensionless velocity fluctuation is defined as
\begin{equation}
  \delta\bm{\varphi}_i = \frac{
    \mathbf{v}_i-\langle\mathbf{v}\rangle
  }{
    \sqrt{\sum_{k=1}^N \|\mathbf{v}_k-\langle\mathbf{v}\rangle\|^2 / N}
  },
  \label{dphi}
\end{equation}
where $\langle \mathbf{v} \rangle = \sum_{i=1}^N \mathbf{v}_i/N$ is the average
velocity.  The connected correlation function is then given by
\begin{equation}
  C(r) = \frac{
    \sum_{i \neq j}
    \delta\bm{\varphi}_i \cdot \delta\bm{\varphi}_j
    \,
    \delta(r-r_{ij})}
  {\sum_{i \neq j}\delta(r-r_{ij})},
  \label{eq:correlation}
\end{equation}
where $r_{ij}=\|\mathbf{r}_i-\mathbf{r}_j\|$ is the distance between agents $i$ and
$j$, and $\delta(r)$ the Dirac delta distribution.
For finite-size systems, one can use the integral of $C(r)$ up to its first zero as an estimation for the susceptibility of the system \cite{attanasi14:_collec_behav_collec_order_wild_swarm_midges},
\begin{equation}
\chi = \max_{r_0}\left(\int_{r<r_0}\!\!\!\!\!\!C(r)d\mathbf{r}\right) .
\label{eq:susceptibility}
\end{equation}
Using Eq.~\eqref{eq:correlation} to compute the correlation requires of a finite value of noise $\eta$ to avoid indeterminate forms.
In order to study the responsiveness of swarms with high degree of alignment, which is guaranteed in the absence of noise, we have set the noise to $\eta=0.04$ and checked that the results are consistent with vanishing noise (see Supplementary Note~IB).

To obtain numerical values of the correlation function $C$ and susceptibility
$\chi$, we compute the histogram of the correlations in the system every
$5\times10^3$ iterations during $2\times10^6$ iterations, after discarding the
first $5\times10^4$ iterations as transient dynamics.  The correlation $C(r)$
shown in Fig.~\ref{correlation_topo} is the average over $400$ histograms
obtained with this procedure.

\subsection{Predator attack}

%
The predator is introduced as an agent that does not participate in the
consensus protocol.  Instead, it is afforded predatory capabilities: it moves
$40\,\%$ faster than swarming agents, systematically in the direction pointing
to the closest one.  When the predator ``catches'' an agent, the latter is
removed from the simulation.  An agent can only detect the presence of the
threat when it is located at a distance smaller than a fixed
``danger-detection'' radius $R_D$; as soon as the agent detects it, an evasive
maneuver is initiated with the agent moving away in the direction opposite to
the predator.
We have set $R_D$ to be constant throughout the simulations and
independent of the sociality between agents.
The fleeing behavior takes
precedence over the collective motion of a particular agent for as long as the
predator lies inside its danger-detection area.  Thus, the agents in this
simulation follow the equations of motion~\eqref{vicsek} with the exception
that
\begin{equation}\label{eq:flee}
  \mathbf{v}_i (t)= v_0 \frac{\mathbf{x}_i(t) -\mathbf{x}_P(t)}{\|\mathbf{x}_i(t) -\mathbf{x}_P(t)\|} \quad \mathrm{if}\quad \|\mathbf{x}_i (t)-\mathbf{x}_P(t)\| < R_D ,
\end{equation}
where $\mathbf{x}_P$ denotes the predator's position.

The mean avoidance time shown in Fig.~\ref{survival_rate} is obtained by
computing the swarm dynamics in the presence of a single predator (introduced after a transient of $2,000$ iterations) for $500$ different runs of $5,000$ iterations each.  The reason for computing several runs instead of running the calculation for longer times is that the results
depend on the density of agents in the swarm, and the repeated removal of
agents by the predator can cause significant changes in the density after long
times.

\subsection{Collective decision-making}

%
The linear threshold model is a generalization of the simple majority vote
model~\cite{aldana04:_phase} where the state of each agent or node $i$ is
determined by a binary variable $s_i=\{0,1\}$.  The dynamics of the model
dictates that, at a given timestep $t$, $s_i(t)$ takes the value $0$ or $1$
according to
\begin{equation}
  s_i(t+1) = \left\{ 
    \begin{array}{lr}
      1 & \mathrm{if }\, \langle s_j(t) \rangle_{j\sim i}  > \theta \\
      0 & \mathrm{otherwise , }
    \end{array}
  \right.
  \label{linear_threshold_protocol}
\end{equation} 
where $\langle \cdot \rangle_{j\sim i}$ is the average over all neighbors of
$i$ and $\theta$ is a parameter that determines the minimum ratio of neighbors
that need to be in the state $s=1$ for an agent to switch to it.

To study the effects of a perturbation on the collective decision-making
process, we consider the following scenario: a given set of $N=2,048$
networked agents reside in the ``ground'' state $s_i=0\,\,\forall i$ when, at
$t=0$, an unspecified perturbation induces a small fraction of ``informed''
agents to abruptly switch to (and remain in) the state $s_{\{j\}}=1$.  This
change propagates through the network and causes more agents to switch from
state $0$ to $1$. If the fraction of initially informed agents is large enough
and the network is connected, the mean polarization $P(t)=\langle s_i(t)
\rangle $ will eventually reach $P=1$. 
One can characterize the  responsiveness of the decision-making process by the speed at which this change propagates through the system, measured by the rate of change in polarization,
\begin{equation}\label{eq:c}
  c = \frac{dP}{dt}=\frac{1}{N}\frac{d}{dt}\sum_{i=1}^N s_i(t) .
\end{equation}

Further details can be found in Supplementary Notes V--VII.

\subsection{Distributed consensus in multi-agent systems}

%
Let us consider a group of $N+1$ identical agents performing a distributed
consensus protocol on their scalar state-variable $x_i(t)$, through a
connected and undirected network.  The system is characterized by the global
state vector $\mathbf{X}(t) = \{x_i(t);i=0,\cdots,N\}$ and the adjacency
matrix of the underlying graph ${\mathbf A}=\{a_{ij};i,j=0,\cdots,N\}$, where
$a_{ij}=1$ if agent $i$ is connected to $j$ and 0 otherwise.  Given a certain connectivity graph, the state of the system evolves according to
\begin{align}
  \frac{dx_i}{dt} &= \frac{\omega_0}{k_i}\sum_{j=0}^N a_{ij} \big( x_j(t) -
  x_i(t) \big) ,   \nonumber \\
  &= \sum_{j=0}^N w_{ij} x_j(t) ,
  \label{eq:local-consensus}
\end{align}
where $\omega_0$ is the natural response frequency of our identical agents,
and $k_i = \sum_{j=0}^N a_{ij}$ is the degree of agent $i$, i.e. its number of
neighbors in the network sense. The quantity $w_{ij}=\omega_0
(a_{ij}/k_i-\delta_{ij})$ is introduced for the sake of a compact notation for
the governing dynamical equations. As is classical with many swarming systems,
these dynamics involve relative output information of
neighboring agents~\cite{olfati-saber07:_consen_cooper_networ_multi_agent_system}.

We model the process of distributed transfer of social information by
considering a leader-follower consensus dynamics.  This is implemented by
affording one agent---say agent $i=0$---with a dynamics not abiding by
Eq.~\eqref{eq:local-consensus}, but instead following an arbitrary
trajectory $x_0(t) = u(t)$. This single control input has a direct effect onto
the dynamics of its $k_0$ neighboring agents, but also has indirect effects
onto the dynamics of many more agents through the coupled set of dynamical
equations~\eqref{eq:local-consensus}. In the presence of this single leader,
Eq.~\eqref{eq:local-consensus} can be recast as
\begin{equation}\label{eq:leader}
  \frac{d x_i}{dt} = \sum_{j=1}^N w_{ij} x_j(t) + w_{i0} u(t) ,
\end{equation}
for $i=1,\cdots,N$.

Despite the static nature of the topology
of interaction, this leader-follower consensus model is a good idealization of
the process of social information transfer occurring in startled schools
of fish or flocks of birds, where one individual has access to privileged
information about a potential threat or other kind of external perturbation.
This temporary leader triggers a wave of agitation that propagates strikingly
fast through the swarm~\cite{herbert-read15:_initiat,%
  hemelrijk15:_scale_free_correl_influen_neigh}. Such waves of agitation are
initiated by extremely rapid changes in the leading agent's state, which very
effectively propagate to all other swarming agents~\cite{attanasi14:_infor}.

Within this leader-follower scheme, one can characterize the responsiveness of
the multi-agent system undergoing the distributed consensus process as its
capacity to follow fast changes in the leader's dynamics,
$u(t)$. Specifically, with an input signal oscillating at the frequency
$\omega$, $u(t)=u_0e^{i\omega t}$, the state of all agents at long times
becomes proportional to $u(t)$ with a factor given by the transfer function, 
\begin{equation}\label{eq:gain} {\mathbf H}(\omega) =
  \lim_{t\rightarrow\infty}\frac{{\mathbf X}(t)}{u(t)} =
  (i\omega\mathbf{I}-{\mathbf W})^{-1}{\mathbf W}_l.
\end{equation}
where $\mathbf{I}$ is the identity matrix of dimension $N$, ${\mathbf W}=\{w_{ij}\}$ is the $N\times N$ consensus protocol matrix and ${\mathbf W}_l=\{w_{i0}\}$ is the $N$-vector resulting from projecting ${\mathbf W}$ onto the subspace of the leader.
This allows us to define the system's responsiveness as the norm of this transfer function, $H^2 =\|{\mathbf H}\|^2=\sum_i |h_{i}(\omega)|^2 $, with $|h_{i}(\omega)|\le1$ for all $i$ and $\omega$~\cite{ogata10:_moder_contr_engin}.
As is clear from Eq.~\eqref{eq:gain}, the gain functions have a nontrivial dependency on the topology of the agents' connectivity, including that of the leading agent, through the entries of $\mathbf{W}$ and $\mathbf{W}_l$.

\section{Acknowledgments}
This work was supported by a grant from the Temasek Lab (TL@SUTD) under the
STARS project (D.M.) and a grant \#SMIG14006 from the Singapore--MIT Alliance
for Research and Technology (SMART) (R.B.).

The authors thank Nikolaj S{\o}rensen for fruitful discussions on state-space
modeling and Dr. Mardavij Roozbehani for his input on the importance of nodal
dynamics for the time-constrained controllability of a system.

\section{Author contributions}
D.M. and R.B. designed the study.
D.M. performed research, developed the analytical and numerical tools.
Y.K.K. performed research and computed the control-theoretic aspects.
R.B. coordinated the study.
D.M. and R.B. analyzed the results and wrote the paper.


\end{document}